\documentclass[twocolumn]{aastex63}

\newcommand{\lya}{Ly$\alpha$}
\newcommand{\ha}{H$\alpha$}
\newcommand{\ergs}{\rm erg\ s^{-1}}

\graphicspath{{./}{figures/}}

\accepted{\today}
\submitjournal{ApJ Letters}

\shorttitle{X-rays from Nearby \lya\ Emitters}
\shortauthors{Dittenber et al.}

\begin{document}

\title{Accretion-Driven Sources in Spatially Resolved Ly$\alpha$ Emitters}

\author{Benjamin Dittenber}
\affiliation{University of Michigan,
Department of Astronomy,
Ann Arbor, MI 48103}

\author{M. S. Oey}
\affiliation{University of Michigan,
Department of Astronomy,
Ann Arbor, MI 48103}

\author{Edmund Hodges-Kluck}
\affiliation{University of Maryland,
Department of Astronomy,
College Park, MD 20742}
\affiliation{X-ray Laboratory,
Code 662, NASA/GSFC,
Greenbelt, MD 20770}

\author{Elena Gallo}
\affiliation{University of Michigan,
Department of Astronomy,
Ann Arbor, MI 48103}

\author{Matthew Hayes}
\affiliation{Stockholm University,
Department of Astronomy 
and Oskar Klein Centre for Cosmoparticle Physics, AlbaNova University Centre, SE-10691,
Stockholm, Sweden}
\author{G\"oran \"Ostlin}
\affiliation{Stockholm University,
Department of Astronomy 
and Oskar Klein Centre for Cosmoparticle Physics, AlbaNova University Centre, SE-10691,
Stockholm, Sweden}
\author{Jens Melinder}
\affiliation{Stockholm University,
Department of Astronomy 
and Oskar Klein Centre for Cosmoparticle Physics, AlbaNova University Centre, SE-10691,
Stockholm, Sweden}




\begin{abstract}
\lya\ emission is a standard tracer of starburst galaxies at high redshift. However, a number of local \lya\ emitters (LAEs) are X-ray sources, suggesting a possible origin of \lya\ photons other than young, hot stars, and which may be active at much later ages relative to the parent starburst.
Resolved, nearby LAEs offer the opportunity to discriminate between diffuse X-ray emission arising from supernova-heated gas, high-mass X-ray binaries (HMXBs), or low-luminosity active galactic nuclei (LLAGN).  We examine archival X-ray imaging from {\sl Chandra} and {\sl XMM-Newton} for 11 galaxies with spatially resolved \lya\ imaging to determine the luminosity, morphology, and spectral hardness of the X-ray sources.  The data are consistent with 9 of the 12, bright \lya\ sources being driven by luminous, $10^{40}\ \ergs$ X-ray sources. 
Half of the 8 {\sl Chandra} sources are unresolved.  The data suggest that nuclear activity, whether from LLAGN or nuclear starbursts, may play an important role in \lya\ emission.  Our results also suggest a significant link between \lya\ emission and HMXBs, ULXs, and/or LLAGN, which would imply that \lya\ may be generated over timescales 1 -- 2 orders of magnitude longer than produced by photoionization from OB stars.  This 
highlights a critical need to quantify the relative contributions of different sources across cosmic time, to interpret \lya\ observations and the resulting properties of distant galaxies.  

\end{abstract}



\section{Introduction} \label{sec:intro}
Studies of galaxies at high redshift often rely on Ly$\alpha$ emission, which is produced in the recombination of ionized hydrogen and predicted to be associated with young massive stellar populations. More than five decades ago,  Ly$\alpha$ was put forth as a tool to study high redshift, star-forming galaxies due to its high luminosity and accessibility from earth-based observatories \citep{partrdige67}. Over the years, it has been used to probe the epoch of reionization \citep[e.g.,][]{malhotra04,kashikawa06,stark17}, and numerous high redshift galaxies have been discovered by narrow-band \lya\ imaging \citep[e.g.,][]{hu96,mallery12} and IFU observations \citep[e.g.,][]{herenz19, hashimoto17}. 

The complexity of Ly$\alpha$ radiative transfer complicates our interpretation of the conditions allowing its escape
from the local environment. 
Since Ly$\alpha$ is a resonant transition, it scatters both spatially and spectrally. In addition,
Ly$\alpha$ emission cannot be spatially resolved at high redshift, and therefore the contributing sources cannot be clearly determined. Thus, often only global measurements of total \lya\ emission from distant galaxies are possible. These factors limit our understanding of the underlying astrophysical processes that generate \lya\ emitters (LAEs) and subsequent interpretation of high redshift observations.  

\lya\ emission is usually produced by photoionization, which in high-redshift, starburst galaxies is often assumed to originate from star-forming H{\sc ii} regions.  However,
accreting compact objects also produce ultraviolet ionizing photons, which therefore are also a potential source of Ly$\alpha$ emission; for example, active galactic nuclei (AGN) are well-known LAEs \citep{calhau20}.  
These alternative sources are important because high-mass X-ray binaries (HMXBs) may considerably extend the period of \lya\ emission from starbursts, since HMXBs form after the most massive stars have expired, and nuclear starbursts may also trigger accretion onto nuclear, massive black holes, which can be sustained at a low level for times on order $10^8$ yr or more. 

Local, spatially resolved LAEs can help clarify the origin of \lya\ emission in starburst galaxies.  Studying LAEs in X-rays can show whether most starburst LAEs are indeed generated by the photoionization from massive stars rather than AGN.  
For example, the starburst interacting galaxy system Haro 11 is an LAE with a bright, hard, compact X-ray source coinciding with the galaxy's only strong \lya\ source. This X-ray emission appears to be due to an ultra-luminous X-ray source \citep[ULX;][]{prestwich15}. 
Sources like this cannot be attributed to shock-heated, diffuse gas from massive-star feedback, and they are more likely due to X-ray binaries (HMXBs) or low luminosity active galactic nuclei \citep[LLAGN;][]{prestwich15, oskinova19}. 

In this work, we examine the X-ray emission from local starburst galaxies with confirmed, resolved \lya\ imaging, and that also have archive X-ray data \citep[e.g.,][]{basu-zych13, brorby16} available.
From these data we can determine whether HMXBs and/or LLAGN play a significant role in powering LAEs and/or facilitating \lya\ escape.
If HMXBs are responsible for LAEs, then Ly$\alpha$ may be still be linked to massive stellar clusters, but at a more evolved stage 
than photoionization by stellar radiation itself.  This would imply that \lya\ emission from starbursts is longer lived than when assumed to originate from only stellar photoionization.

\section{Sample and Data Analysis}
We use the local sample of starburst galaxies from \citet{ostlin09} and the Lyman Alpha Reference Sample \citep[LARS;][]{ostlin14}, which are good analogs to high redshift LAEs.  These two samples target galaxies that are at distances of roughly 40 to 250 Mpc, and all have uniform, spatially resolved, \lya\ imaging from the {\sl Hubble Space Telescope}. 
The LARS objects were selected to omit targets with any nebular evidence of AGN \citep{ostlin14}. In particular, the line width of H$\alpha$ must have FWHM $<300$ km/s, and second, the galaxies have H{\sc ii}-region-like, optical emission-line ratios. 

\begin{figure*}
\gridline{
\fig{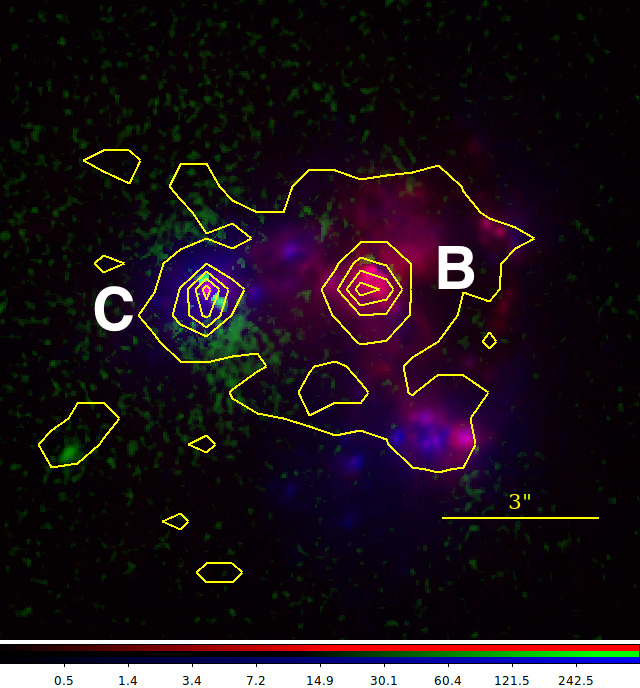}{0.3\textwidth}{(a) Haro 11}
\fig{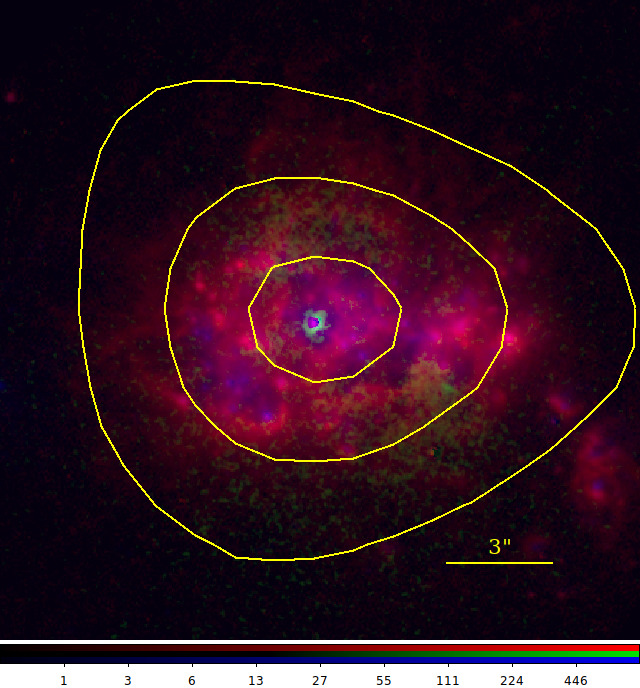}{0.3\textwidth}{(b) ESO 338-IG004}
\fig{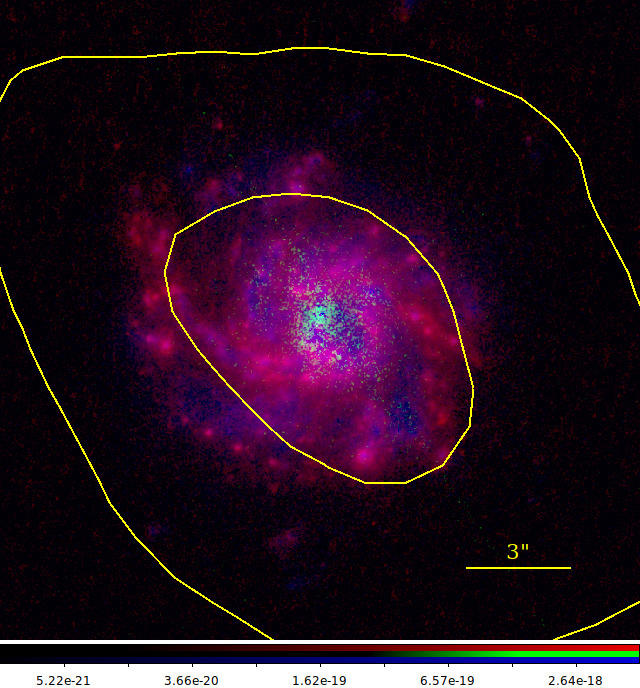}{0.3\textwidth}{(c) IRAS 08339$+$6517}
}
\gridline{
\fig{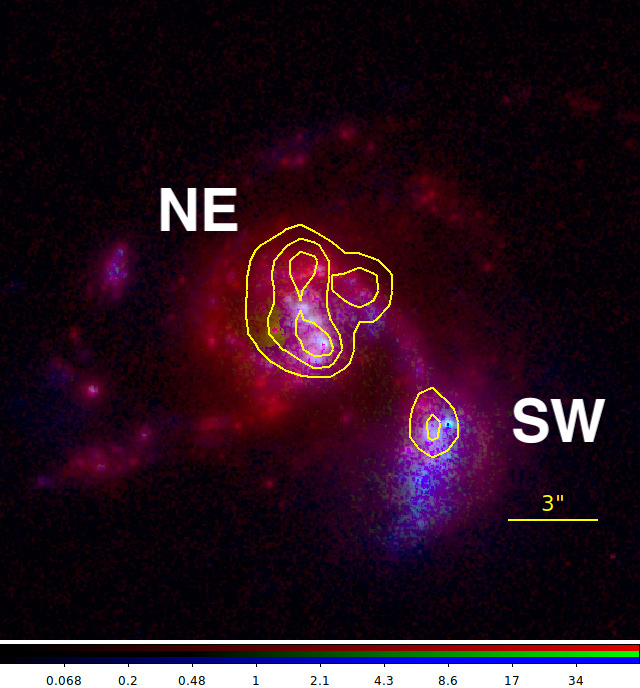}{0.3\textwidth}{(d) NGC 6090}
\fig{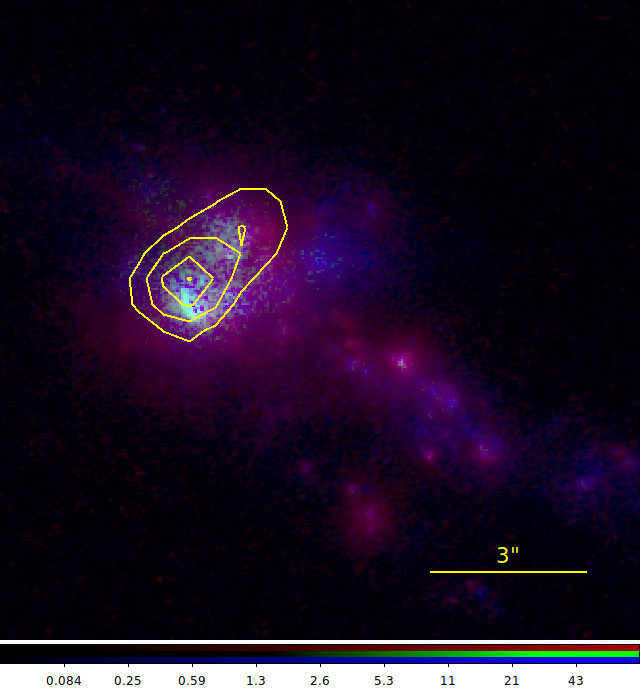}{0.3\textwidth}{(e) LARS 01}
\fig{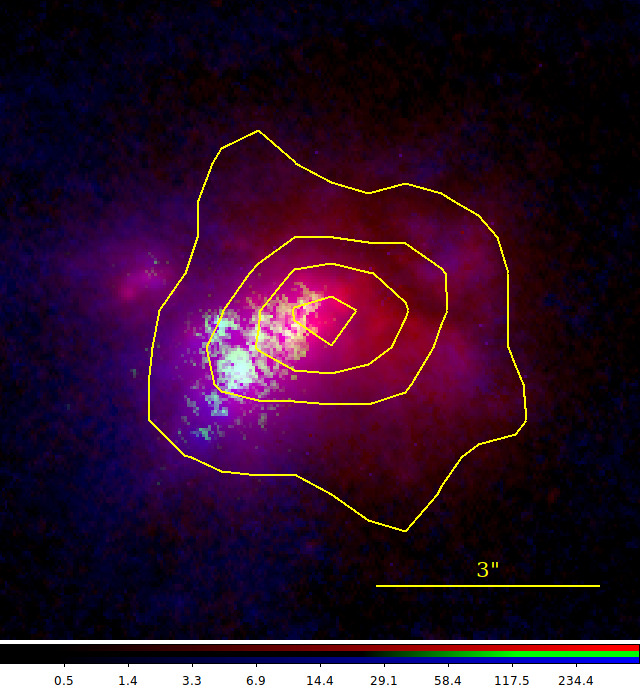}{0.3\textwidth}{(f) LARS 03}
}
\caption{Composite RGB images of galaxies in our sample. H$\alpha$,
  continuum-subtracted Ly$\alpha$, and FUV are shown by red, green,
  and blue, respectively.  Contours represent 0.3--10~keV X-ray surface
  brightness with the significance for each object (in units of
  $\sigma$, where $\sigma$ is measured from a local background
  annulus) and Gaussian smoothing kernel in parentheses:  
Haro~11: 1.7, 15, 28, 42, and
  75$\sigma$ (no smoothing); ESO~338-IG004: 2.0, 6.0, and 11$\sigma$
  (2 pixels); IRAS~08339$+$6517: 5.9 and 14$\sigma$ (2 pixels);
  NGC~6090: 3.9, 5.1, and 5.9$\sigma$ (2 pixels); LARS~01: 4.0, 6.0,
  and 8.0$\sigma$ (3 pixels); LARS~03: 2.8, 4.1, 4.9, and 6.3$\sigma$
  (3 pixels).
  \label{f_img1}}
\end{figure*}

\begin{figure*}[hbt!]
\gridline{
\fig{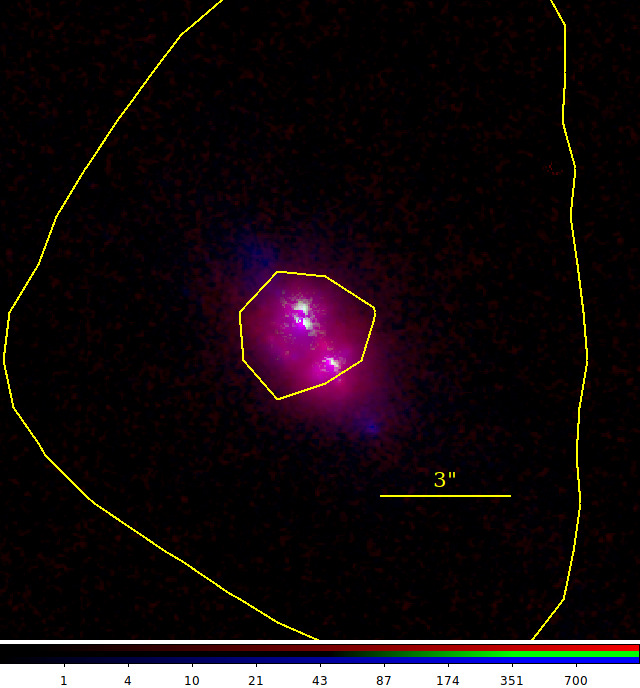}{0.3\textwidth}{(g) LARS 07}
\fig{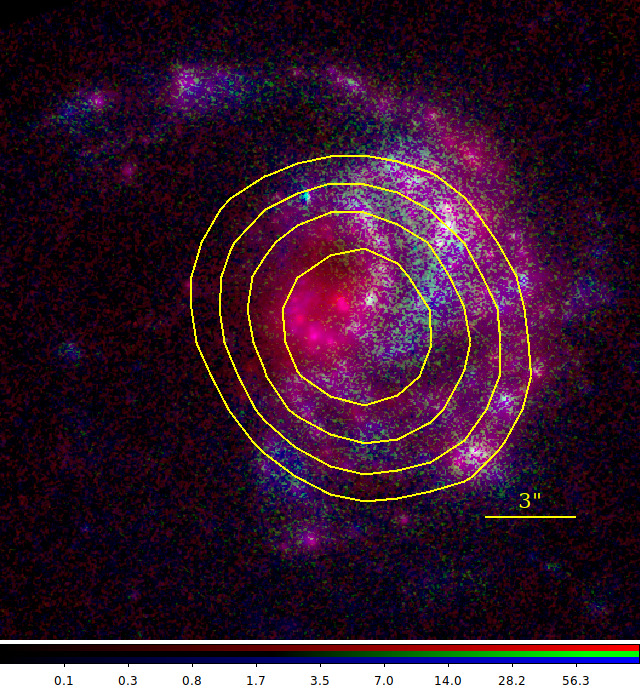}{0.3\textwidth}{(h) LARS 08}
\fig{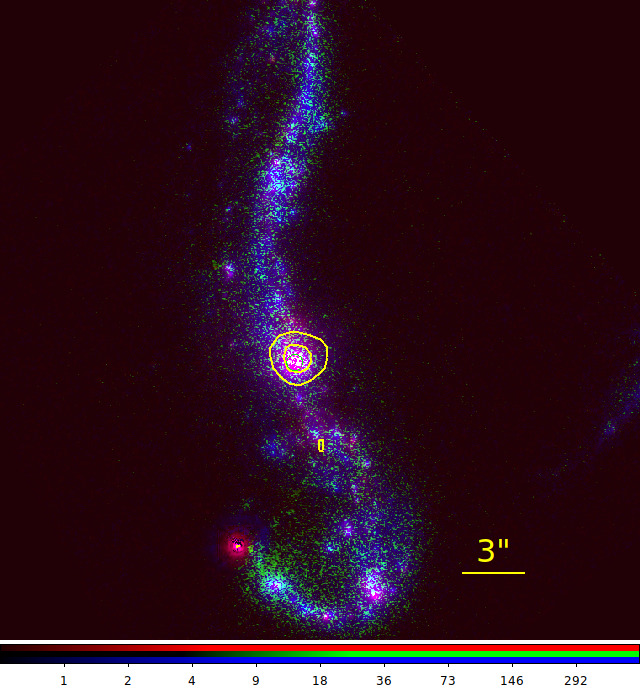}{0.3\textwidth}{(i) LARS 09}
}\gridline{
\fig{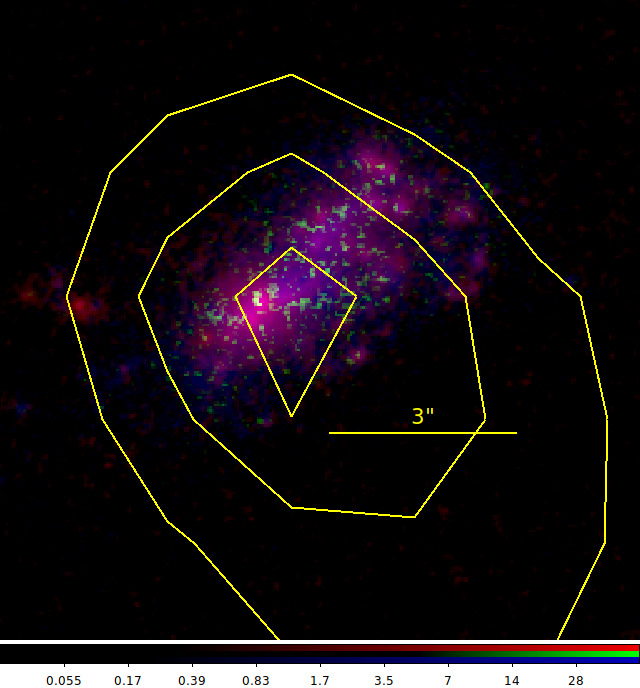}{0.3\textwidth}{(j) LARS 10}
\fig{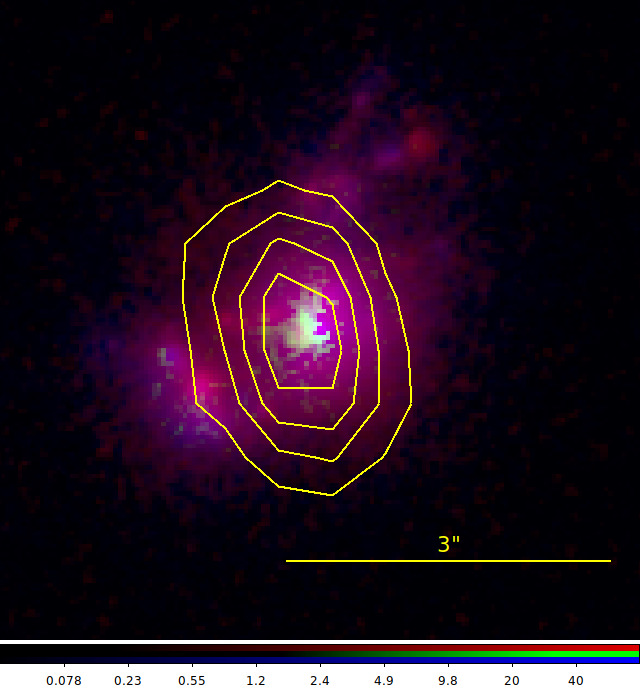}{0.3\textwidth}{(k) LARS 12}
}
\caption{Same as Figure~\ref{f_img1} for the remainder of our sample.
LARS~07: 1.5 and 3.1$\sigma$ (3 pixels); LARS~08: 4.8, 5.8, 6.8, and 7.8$\sigma$ (3 pixels); LARS~09: 1.8 and 3.5$\sigma$ (3 pixels); LARS~10: 1.1, 2.2, and 3.2$\sigma$ (3 pixels); LARS~12: 1.6, 3.3, 5.0, and 6.6$\sigma$ (3 pixels). 
  \label{f_img2}}
\end{figure*}

From these samples, we select the objects that are confirmed LAEs and also have publicly available archival X-ray data from the {\sl Chandra X-ray Observatory} ACIS-S and {\sl XMM-Newton} EPIC-pn and MOS instruments.
This yields 11 galaxies, given in Table~\ref{table:obs}, and shown in three-color images in Figures~\ref{f_img1} and \ref{f_img2}. 

We followed standard analysis procedures, which for \textit{Chandra} involved using the 
CIAO v4.9 software package.  We used the {\tt chandra\_repro} script to create level=2 event files and {\tt deflare} to remove periods of high background flaring, excluding periods where the chip count rate exceeds the average by more than 3$\sigma$. We identified sources using the {\tt wavdetect} tool with default detection thresholds on the 0.3-10~keV band, then calibrated the astrometry by cross-matching at least three detected sources with the Sloan Digital Sky Survey (SDSS) catalog or the USNO~B1.0 catalog.  We obtained the X-ray fluxes with the CIAO {\tt srcflux} tool, using the Ideal PSF method for a circular aperture encompassing the entire galaxy's emission, centered on the source, and using a source-free local background annulus. While this represents the integrated emission from all sources, there was one dominant source in each galaxy, except for the merging components of Haro~11 and NGC 6090, which were treated as individual galaxies.
For the model flux, we assumed an absorbed power-law spectrum with a photon index $\Gamma=2.0$ and with Galactic absorption based on the {\tt colden} tool. We caution that $\Gamma$ has some variation for HMXBs and LLAGN \citep[e.g.,][]{terashima2002, sazonov2017}.  For $\Gamma=(1.5, 2.5),$ the resulting flux is (0.8, 3) times the value obtained for $\Gamma=2.0$.  For any objects dominated by soft sources having larger values of $\Gamma$, the flux could be further underestimated.

For \textit{XMM} data, we followed a similar procedure, using the
SAS v17.0.0 software, and using the {\tt epchain} and {\tt emchain} scripts to calibrate the data and produce level=2 event files. We filtered on the background light curve in the same way as for the \textit{Chandra} data, and then detected sources and measured fluxes using the {\tt edetect\_chain} script.  We set the image {\tt binsize} to 22, corresponding to a resolution of 1.1$\arcsec$, and processed the separate images for 0.5-8.0 keV.  EPIC-pn vs EPIC-MOS observations were chosen based on S/N.
The 90\% confidence X-ray fluxes are given in Table~\ref{chartable}. 

\begin{deluxetable*}{llccllc}
\tablenum{1}
\tabletypesize{\scriptsize}
\tablecaption{\label{table:obs} X-ray Observations}
\tablewidth{0pt}
\tablehead{
\colhead{Galaxy} & \colhead{Alt. ID} & \colhead{R.A.} & \colhead{Dec} &  \colhead{Observatory} & \colhead{ObsID} & \colhead{Exposure} \\
 & & \colhead{(J2000)} & \colhead{(J2000)} & &  & \colhead{(ks)}
}
\startdata
Haro 11         & ESO 350-IG038            & 00:36:52.70 & $-$33:33:17.0  & Chandra    & 16695  & 24.74 \\
NGC 6090        & Mrk 496                   & 16:11:40.70 & $+$52:27:24.0  & Chandra   & 6859  & 14.79 \\
IRAS 08339+6517 &                           & 08:38:23.18 & $+$65:07:15.2  & XMM-Newton  & 0111400101  & 57.65 \\
ESO 338-IG004  & Tol 1924--416              & 19:27:58.17 & $-$41:34:32.2  & XMM-Newton  & 0780790201 & 24 \\
LARS 01        &                           & 13:28:44.05 & $+$43:55:50.5 & Chandra & 19442  & 33.41 \\
LARS 03         & UGC 08335                 & 13:15:34.98 & $+$62:07:28.7  & Chandra   & 7810  & 14.85 \\
LARS 07         & Ton 151                   & 13:16:03.92 & $+$29:22:54.1  & XMM-Newton   & 0780790401  & 22 \\
LARS 08         & WISEA J125013.82+073444.7 & 12:50:13.85 & $+$07:34:44.5  & XMM-Newton   & 0780790101  & 20 \\
LARS 09         & KUG 0820+282              & 08:23:54.95 & $+$28:06:21.6  & Chandra    & 13012  & 8.89 \\
LARS 10         & Mrk 61                    & 13:01:41.53 & $+$29:22:52.9  & Chandra    & 15065  & 14.87 \\
LARS 12         & SBS 0934+546              & 09:38:13.50 & $+$54:28:25.1  & Chandra   &  16018  & 15.47
\enddata
\end{deluxetable*}

\setcounter{table}{1}
\begin{longrotatetable}
\begin{deluxetable*}{lcccccccc}
\rotate
\tablecaption{X-ray Data for Local \lya-emitting Galaxies\label{chartable}}
\tabletypesize{\scriptsize}
\tablehead{
\colhead{Name} & \colhead{Instrument} & 
\colhead{FWHM\tablenotemark{a}} & \colhead{$f_X (0.5 - 8$ keV} & 
\colhead{$L_X (0.5 - 8)$ keV} & \colhead{HR} & 
\colhead{SFR\tablenotemark{b}} & \colhead{Distance\tablenotemark{c}} & 
\colhead{$12 + \log$(O/H)\tablenotemark{d}} \\
\colhead{} & \colhead{} &
\colhead{(arcsec)} & \colhead{($10^{-14}\ \rm erg\ s^{-1} cm^{-2}$)} & \colhead{(10$^{40}$ erg s$^{-1}$)} & \colhead{} & 
\colhead{(M$_\odot$ yr$^{-1}$)} & \colhead{(Mpc)} & \colhead{}
} 
\startdata
Haro 11 & ACIS-S & \nodata & $21.2_{-1.5}^{+1.5}$ & $19.3_{-1.4}^{+1.4}$ & \nodata & 5.15 & $87\pm6.1$ & 7.9 \\
\hspace{3mm}(Knot B)\tablenotemark{e} & ACIS-S & 0.77 & $11.8_{-1.1}^{+1.1}$ & $10.7_{-1.0}^{+1.0}$ & $0.40_{-0.32}^{+0.39}$ & 0.86 & $87\pm6$ & 8.3 \\
\hspace{3mm} Knot C & ACIS-S & {$\bf \leq 0.5$} & $7.76_{-0.91}^{+0.91}$ & $7.07_{-0.83}^{+0.83}$ & \nodata & 0.09 & $87\pm6$ & 7.8 \\
NGC 6090 & ACIS-S & \nodata & $14.2_{-1.3}^{+1.3}$ & $28.7_{-2.6}^{+2.6}$ & \nodata & 4.75 & $130\pm9$ & 8.8  \\
\hspace{3mm}SW Galaxy & ACIS-S & {$\bf \leq 0.5$} & $2.98_{-0.61}^{+0.62}$ & $6.02_{-1.23}^{+1.25}$ & \nodata & 1.57 & $130\pm9$ & 8.8 \\
\hspace{3mm}NE Galaxy\tablenotemark{g} & ACIS-S & 2.4 & $10.4_{-1.1}^{+1.1}$ & $21.0_{-2.2}^{+2.2}$ & $-0.21_{-0.16}^{+0.18}$ & 3.18 & $130\pm9$ & 8.8 \\
IRAS 08339$+$6517\tablenotemark{f} & EPIC-MOS & 10 & $41.5_{-1.7}^{+1.7}$ & $36.5_{-1.5}^{+1.5}$ & $0.02\pm0.05$ & 7.86 & $86\pm6$ & 8.7 \\
ESO 338-IG004 & EPIC-MOS & {\bf 4.5} & $35.8_{-1.8}^{+1.8}$ & $6.84_{-0.34}^{+0.34}$ & $0.27\pm0.05$ & 3.90 & $40\pm3$ & 7.9 \\
LARS 01 & ACIS-S & {$\bf \leq 0.5$} & $1.96_{-0.43}^{+0.44}$ & $3.75_{-0.82}^{+0.84}$ &  $0.04_{-0.27}^{+0.39}$ & 6.52 & $126\pm9$ & 8.3 \\
LARS 03 & ACIS-S & 1.1 &  $15.2_{-1.4}^{+1.4}$ & $35.1_{-3.2}^{+3.2}$ & $0.33_{-0.11}^{+0.16}$ & 26.3 & $139\pm10$ & 8.4 \\
LARS 07 & EPIC-pn & {$\bf \leq 12.5$} & $1.00_{-0.20}^{+0.20}$ & $3.49_{-0.70}^{+0.70}$ & $-0.24\pm0.25$ & 9.27 & $171\pm12$ & 8.4 \\
LARS 08 & EPIC-pn & {$\bf \leq 12.5$} & $8.18_{-0.50}^{+0.50}$ & $29.3_{-1.8}^{+1.8}$ & $-0.11\pm0.08$ & 36.8 & $173\pm12$ & 8.5 \\
LARS 09\tablenotemark{g} & ACIS-S & 0.57 & $4.86_{-1.15}^{+1.15}$ & $25.7_{-6.1}^{+6.1}$ & \nodata & 40.7 & $210\pm15$ & 8.4 \\
LARS 10 & ACIS-S & 1.5 & $0.41_{-0.28}^{+0.43}$ & $3.23_{-2.22}^{+3.39}$ & \nodata &  5.59 & $258\pm18$ & 8.5 \\
LARS 12 & ACIS-S & {$\bf \leq 0.5$} & $1.07_{-0.35}^{+0.45}$ & $26.4_{-8.6}^{+11.1}$ & \nodata &  97.0 & $454\pm32$ & 8.4 \\
\enddata
\tablenotetext{a}{Instrument PSFs:  {\sl XMM} EPIC pn $= 12.5\arcsec$, {\sl XMM} EPIC MOS $= 4.3\arcsec$, {\sl  Chandra} ACIS $ = 0.5\arcsec$.  Boldface values are unresolved.}
\tablenotetext{b}{SFR based on \ha\ compiled by \citet{hayes14} for LARS galaxies, and computed from NUV and FIR fluxes for \citet{ostlin09} galaxies. For ESO 338-IG004 we adopt 2255 \AA\ flux $= 4\times10^{-14}\ \ergs\ \rm cm^{-2}\ \AA^{-1}$ interpolated from adjacent NUV bands in \citet{ostlin09}. Uncertainties are on the order of 0.3 dex or more.} 
\tablenotetext{c}{Distance from NED.}
\tablenotetext{d}{LARS galaxy metallicities compiled by \citet{pardy14}; remaining objects from \"Ostlin et al. (2009)}
\tablenotetext{e}{Haro 11-B is not a significant \lya\ source and is included here for comparison.}
\tablenotetext{f}{Severely off-axis {\sl XMM} observation; may be unresolved.}
\tablenotetext{g}{Multiple X-ray sources. FWHM is measured from the brightest source; other X-ray quantities are integrated over the multiple sources.}
\end{deluxetable*}
\end{longrotatetable}

\section{X-rays from \lya\ Emitters}

We find that in nearly every galaxy, there is X-ray emission consistent with the location of the \lya\ emission identified by \citet{hayes14} and \citet{ostlin09}.  One exception is LARS 09, which has a nuclear X-ray source while the \lya\ source is at large galactocentric radius. Also, Haro~11 has two strong ULXs, while only one is a strong LAE \citep[][]{prestwich15}; and NGC 6090 is a major merger with \lya\ and X-ray emission from both the NE and SW components. The component sources in these mergers are listed separately in Table~\ref{chartable}.

\subsection{Angular Sizes} 
We can use the angular size of the source to constrain the origin of the X-ray emission. Diffuse emission arising from mechanical feedback would appear extended and resolved.  On the other hand, LLAGN should appear as nuclear point sources, so we expect the corresponding X-ray sources to be unresolved.  However, a luminous HMXB may also appear as a single ULX, and in more distant galaxies, multiple HMXBs may also appear unresolved; for the most distant object, LARS 12, the 0.5$\arcsec$ ACIS PSF corresponds to $\sim900$ pc. 

The source sizes were estimated from the FWHM of axisymmetric radial profiles, 
using radial annuli large enough to account for pixelization and to cleanly measure the FWHM.
The on-axis PSF of {\sl Chandra} ACIS is 0.5$\arcsec$. For {\sl XMM-Newton} EPIC-pn, MOS-1 and MOS-2, the FWHM are 12.5$\arcsec$, 4.3$\arcsec$, and 4.4$\arcsec$, respectively.  

We find that Haro~11-C, NGC 6090-SW, LARS 01, LARS 12, and possibly LARS 09, are unresolved point sources, corresponding to at least half of the 8 LAE sources observed with ACIS.
An additional 4 sources are detected only by XMM, which has much larger PSFs that cannot resolve point sources in these galaxies: ESO 338-IG004, LARS 07, and LARS 08 are unresolved, and only IRAS 08339$+$6517 is extended.
Overall, 7 of the 12 sources are technically unresolved (boldface in Table~\ref{chartable}).

\subsection{X-ray Luminosities}

We expect to find HMXBs in these starburst galaxies, since 
there is a direct, empirical relationship between the total X-ray luminosity $L_X$ from HMXBs, and star-formation rate (SFR; Lehmer et al. 2010; Mineo et al. 2012) in star-forming galaxies.  
This relation also depends on metallicity, so that (Brorby et al. 2016):
\begin{eqnarray} \label{eq_brorby}
   \log L_X/({\rm \ergs}) = \log {{\rm SFR/(M_\odot\ yr^{-1})}} + 
   \nonumber \\
   b\ {\rm (12+\log(O/H) - 8.69)} + 39.49
\end{eqnarray}
where $L_X$ is the luminosity in the 0.5 -- 8 keV band, $b = -0.59\pm 0.13$ (Brorby et al. 2016), and solar metallicity corresponds to ${\rm 12+\log(O/H)}= 8.69$.
Low-mass X-ray binaries are not a significant source of X-ray emission in our starbursts, having orders of magnitude lower total $L_X$ \citep{mineo12}.

\begin{figure*}
\centering
    \includegraphics[width=\linewidth]{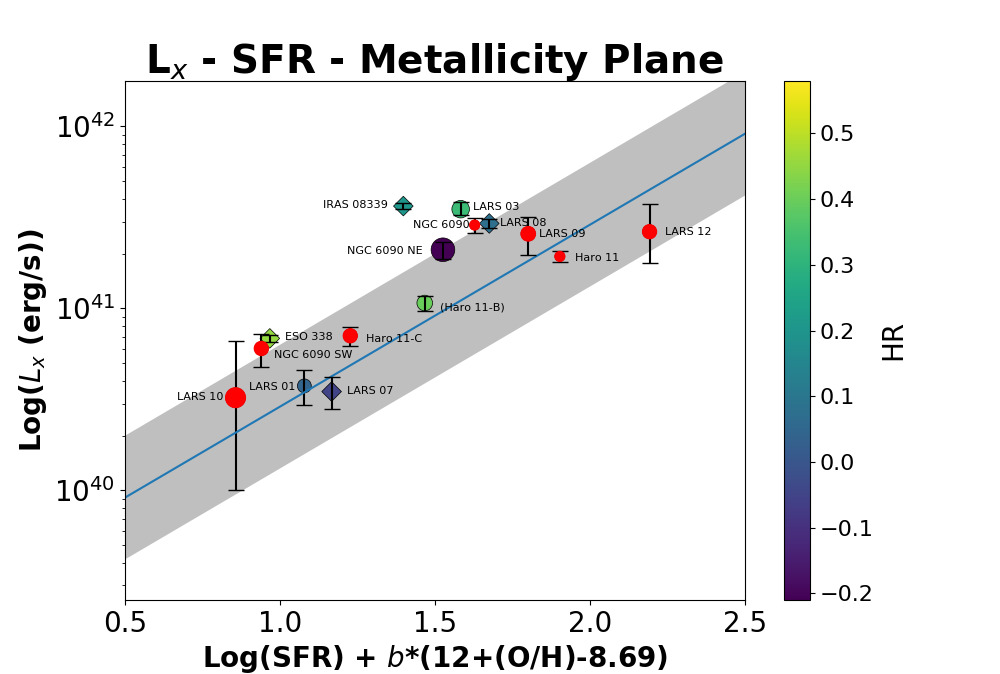}
    \caption{$L_X$ - SFR - metallicity relation for star-forming galaxies, showing our sample and the relation from Brorby et al. (2016; blue line).  Their observed 0.34 dex dispersion in the relation is shown by the grey band.
Hardness ratio is indicated with the shown color scale; red points indicate objects without enough counts for a reliable HR estimate.  Circles show {\sl Chandra} ACIS observations, with point size scaled to source FWHM; diamonds indicate {\sl XMM} data, which cannot resolve point sources in our sample.  Haro 11-B is not a significant LAE, but is included for completeness.  Data for Haro 11 and NGC 6090 are shown for individual sub-regions, as well as for total integrated values.
\label{f_brorby}}
\end{figure*}

If $L_X$ is significantly more luminous than expected from this relation, then an LLAGN may be present. 
In Figure~\ref{f_brorby}, we evaluate this possibility on a statistical basis,
calculating $L_X$ from our measured flux values in the $0.5-8.0$ keV band.  The $L_X$ are in the range $3 - 36 \times10^{40}\ \ergs$ (Table~\ref{chartable}). 
For the LARS galaxies, we derive the SFR from \ha\ luminosities given by \citet{hayes14}.  For the remaining galaxies, we use the relation from 
\citet{mineo12} that sums the SFRs derived from the NUV and FIR luminosities, using the 2255\ \AA\ and FIR fluxes given by \citet[][Table~\ref{chartable}]{ostlin09}.  

We find that the LAE galaxies have large $L_X > 10^{40}\ \ergs$, and generally lie along the $L_X$-SFR-metallicity relationship (Figure~\ref{f_brorby}).  More objects are found above the relation than below.
Equation~\ref{eq_brorby} is based on NUV$+$FIR fluxes \citep{brorby16}, which tend to overestimate SFR relative to \ha\ \citep{hirashita2003};
thus there may be a slight systematic offset between the {\"Ostlin} and LARS subsamples.  However, it is the \"Ostlin galaxies that appear to deviate to higher $L_X$, rather than the LARS galaxies.  
The already high $L_X$ suggests that soft, thermal X-ray sources are unlikely to dominate, since such sources would have underestimated values of $L_X$.

IRAS 08339$+$6517 has the largest excess $L_X$ (Figure~\ref{f_brorby}).
\citet{oti-floranes14} find that its SFR is consistent with their inferred $L_X = 2\times 10^{41}\ \ergs$, which is lower than our result for this galaxy, but the lower $L_X$ would still correspond to a significant excess in Figure~\ref{f_brorby}. \citet{oti-floranes14} estimated the SFR and expected $L_X$ from detailed population synthesis modeling, and so it is hard to compare their conclusion with the empirical $L_X$-SFR relation.  Their analysis may demonstrate the limitations of using this relation for individual galaxies.  On the other hand, this galaxy may host an LLAGN candidate.  

\newpage{}

\subsection{Hardness Ratios}
We compute hardness ratios (HRs) from the model-independent fluxes measured for each source through CIAO's \textbf{srcflux} function:
\begin{equation}
    {\rm HR}_{xy}=\frac{F(x)-F(y)}{F(x)+F(y)}
\end{equation}
where $F(x)$ and $F(y)$ are the model-independent flux for the hard energy band ($2.0-8.0$ keV) and soft band ($0.5-2.0$ keV), respectively.  In some cases, the count rates are too low to obtain a useful HR, and we report in Table~\ref{chartable} only values where the HR errors are $< 0.5$.  Figure~\ref{f_brorby} indicates the hardness ratios by point color.  All of the measured HR for our \lya\ sources are consistent with $\Gamma=2.0\pm0.5$, which corresponds to HR $=0.0\pm0.33$.  The HR may be significantly overestimated if there is a high HI column; however, the existence of \lya\ emission argues against high columns associated with most objects.

We expect LLAGN to show harder states than HMXBs, although there is some variance \citep[e.g.,][]{ho08}. 
Interestingly, while the errors are large, there does appear to be a tendency for the objects with excess $L_X$ to have harder HRs in Figure~\ref{f_brorby}.  

\subsection{Morphology} 

The morphology of these galaxies provides further insight on the origin of their X-ray emission.  
We see that spatially resolved observations are essential, since in at least one galaxy, LARS~09, the \lya\ and X-ray emission are completely unrelated (Figure~\ref{f_img2}c):  the X-ray source is in the nucleus, while the \lya\ is in the southern outskirts of the galaxy.  
In Haro~11, the LAE Haro~11-C does not correspond to the dominant X-ray source, Haro~11-B (Figure~\ref{f_img1}a; Table~\ref{chartable}).  
The former is a ULX, while the latter could be a faint LLAGN ($L_X = 1.1\times 10^{41}\ \ergs$) \citep{prestwich15}.  
Similarly, ESO 338-IG004 has 
a hard HR, consistent with a candidate LLAGN (Table~\ref{chartable}), but \citet{oskinova19} recently showed that this is a non-nuclear X-ray source, which is spatially unrelated to the nuclear LAE.  They suggest that the object is an intermediate-mass black hole (IMBH) candidate.

For NGC 6090-NE, LARS 03, LARS 08, and LARS~10, the \lya\ and X-ray emission are spatially consistent with a physical association but their morphologies do not correlate strongly.  Thus, a causal relation between them is ambiguous.  Except for LARS 08, these galaxies show extended emission in both X-ray and \lya, suggesting a possibly important role for luminous star-forming regions and their feedback (Figures~\ref{f_img1}, \ref{f_img2}).
LARS 08 has marginal $L_X$ excess, and its Ly$\alpha$ is coincident with non-nuclear UV continuum emission.  If the bright, unresolved X-ray source (Figure~\ref{f_img2}b) is however due to an LLAGN candidate, it is likely to be located at the galaxy nucleus, which is heavily obscured by dust \citep{hayes14}.  Thus, while this galaxy may host an LLAGN, it also does not seem likely to be responsible for the \lya\ emission, which is probably due to luminous star-forming regions.  

For most of the remaining galaxies, the X-ray emission is consistent, within the astrometric errors, with originating at a nuclear position, or in the case of major mergers, a position consistent with the nucleus of one of the two merging objects. 
While {\sl XMM} data is spatially unresolved, we note that all but one of the 8 galaxies observed with {\sl Chandra} show nuclear X-ray emission, and in Haro~11-C, NGC 6090-SW, LARS~01, and LARS 12 these nuclear sources are unresolved.
\citet{lehmer10} suggest that nuclear X-ray emission $L_X > 10^{40}\ \ergs$ can usually be attributed to AGN.  In any case,
our findings suggest that X-ray emission from LAEs is often associated with nuclear activity, whether LLAGN candidates, or nuclear starbursts. 

\section{Discussion}

Our sample of starburst galaxies with spatially resolved \lya\ shows that luminous X-ray sources cannot be ruled out as the origin of \lya\ emission in 9 of the 12 \lya\ sources.  In particular,
we detect an X-ray source brighter than ${10^{40}\ \ergs}$ that may be associated with the \lya\ emission in each of the 12 \lya\ sources, except for LARS 08, LARS~09, and ESO 338-IG004.  Of the 9 remaining sources, 7 have {\sl Chandra} observations, and 4 are unresolved X-ray point sources.  Extended X-ray emission also does not rule out HMXBs or LLAGN as \lya\ sources.
Although the selection bias is difficult to evaluate, these results are remarkable, and suggest a significant link between X-rays and observed \lya.  

Extrapolating from $L_X$ down to 13.6~eV yields predicted \lya\ luminosities compatible with those observed,
for a typical power-law index $\Gamma = 1.5-2.0$.
Our findings are consistent with those of \citet{bluem19}, who find that 5 of their 8 blue compact galaxies that are Lyman continuum-emitting candidates show significant X-ray excess; and \citet{svoboda19} also find X-ray excess in 2 of their 3 Green Pea galaxies.
As noted in Section 3.2, our sample similarly shows a tendency toward excess $L_X$.

ULXs and/or LLAGN may play an important role in \lya\ emission.
IRAS 08339$+$6517 has significant excess $L_X$. Also, LARS 03 is elevated in both $L_X$ and HR, although it is extended (Figure~\ref{f_img1}f).
These objects are consistent with recent identification of strong AGN candidates in dwarf galaxies with nebular diagnostics of only star formation \citep[e.g.,][]{baldassare18,reines13}.  
And although LLAGN are specifically excluded from equation~\ref{eq_brorby}, galaxies that lie on this relation may still harbor them, e.g., LLAGN candidate Haro~11-B \citep[][Figure~\ref{f_brorby}]{prestwich15}.

Since all of our objects lie on, or above, the $L_X$ - SFR relation, the possible role of HMXBs in driving the LAEs further suggests that compact objects may be a major, and perhaps even dominant, source of Ly$\alpha$ emission in galaxies. 
\citet{basu-zych13} also find above-average $L_X$ in their sample of 6 Lyman-break analog galaxies, which includes 2 members of our sample.  
Due to \lya\ scattering properties, feedback and outflows are likely important in allowing \lya\ to escape \citep[e.g.,][]{hayes2007,wofford13,orsi12}.  Large numbers of HMXBs form in starbursts after 10 -- 20 Myr, when supernova feedback has had time to clear optically thin pathways, and after ionizing OB stars have expired.  \citet{pakull10} also show that ULXs in star forming galaxies can have mechanical feedback that exceeds their X-ray luminosity by orders of magnitude.  Similarly, LLAGN could both produce Ly$\alpha$ and generate escape avenues via disk winds and jet feedback.  

The possible role of HMXBs and LLAGN in powering \lya\ sources implies that
LAEs may be much longer lived than massive stars in the parent starbursts.  HMXBs dominate populations at ages of 10 -- 20 Myr, and low-level accretion in an LLAGN can be sustained on timescales 10$\times$ longer or more.  This has far-reaching implications for interpreting \lya\ observations and understanding cosmic reionization.  
We also note that the possible presence of LLAGN in many star-forming galaxies could also affect calibration of the $L_X$ - SFR relation.  
Moreover, HMXBs are believed to be responsible for an earlier era of cosmic heating preceding reionization \citep{fragos13}, to be detected with new-generation 21-cm surveys. If LLAGN are confirmed in our sample, it would not necessarily affect the impact of HMXBs on cosmic dawn, but rather highlights the need to clarify the relative contributions of LLAGN, HMXBs, and stellar photoionization to the production of \lya\ over cosmic time \citep{lehmer16}.

\acknowledgments

We thank the anonymous referee for helpful comments.
This work was supported in part by NASA grant 80NSSC18K0376 to E.H.K. and HST-GO-15352.002-A to M.S.O.
M.H. and G.O. acknowledge the support of the Swedish Research Council, Vetenskapsr{\aa}det and the Swedish National Space Agency (SNSA).  M.H. is a
Fellow of the Knut and Alice Wallenberg Foundation.

\bibliographystyle{aasjournal}{}




\end{document}